\documentclass[11pt]{article}
\usepackage{amssymb}
\usepackage{amsmath}
\usepackage[all]{xy}
\usepackage{times}
\usepackage{graphicx}

\title{Descriptive Set Theory and  $\omega$-Powers of Finitary  Languages}
\author{Olivier FINKEL and Dominique LECOMTE$^1$}
\date{March 18, 2020}

\def\ufootnote#1{\let\savedthfn\thefootnote\let\thefootnote\relax
\footnote{#1}\let\thefootnote\savedthfn\addtocounter{footnote}{-1}}

\newcommand{\fa}{\forall}
\newcommand{\Ga}{\Gamma}
\newcommand{\Gas}{\Gamma^{<\om}}
\newcommand{\Si}{\Sigma}
\newcommand{\Sis}{\Sigma^{<\om}}
\newcommand{\ra}{\rightarrow}

\newcommand{\la}{language}
\newcommand{\ite}{\item}

\newcommand{\ol}{ $\omega$-language}

\newcommand{\om}{\omega}

\newcommand{\noi}{\noindent}
\newcommand{\tla}{\twoheadleftarrow}
\newcommand{\de}{deterministic }
\newcommand{\proo}{\noi {\bf Proof.} }

\newcommand{\Sio}{\Si^\om}

\newcommand{\vp}{\varphi}

\newcommand{\borom}{{\bf\Delta}^{0}_{\omega}}
\newcommand{\borxi}{{\bf\Delta}^{0}_{\xi}}
\newcommand{\bormxi}{{\bf\Pi}^{0}_{\xi}}

\newcommand{\bormone}{{\bf\Pi}^{0}_{1}}
\newcommand{\ca}{{\bf\Pi}^{1}_{1}}
\newcommand{\bormtwo}{{\bf\Pi}^{0}_{2}}

\newcommand{\bormom}{{\bf\Pi}^{0}_{\omega}}

\newcommand{\borel}{{\bf\Delta}^{1}_{1}}
\newcommand{\Borel}{{\it\Delta}^{1}_{1}}
\newcommand{\borone}{{\bf\Delta}^{0}_{1}}
\newcommand{\bortwo}{{\bf\Delta}^{0}_{2}}

\newcommand{\boraone}{{\bf\Sigma}^{0}_{1}}
\newcommand{\boratwo}{{\bf\Sigma}^{0}_{2}}

\newcommand{\boraom}{{\bf\Sigma}^{0}_{\omega}}
\newcommand{\boraxi}{{\bf\Sigma}^{0}_{\xi}}
\newcommand{\ana}{{\bf\Sigma}^{1}_{1}}

\newcommand{\Ana}{{\it\Sigma}^{1}_{1}}
\newcommand{\Boraone}{{\it\Sigma}^{0}_{1}}
\newcommand{\Borone}{{\it\Delta}^{0}_{1}}
\newcommand{\Bormone}{{\it\Pi}^{0}_{1}}

\newcommand{\Ca}{{\it\Pi}^{1}_{1}}

\newcommand{\Boran}{{\it\Sigma}^{0}_{n}}
\newcommand{\Bormn}{{\it\Pi}^{0}_{n}}
\newcommand{\Born}{{\it\Delta}^{0}_{n}}
\newcommand{\Boranpo}{{\it\Sigma}^{0}_{n+1}}
\newcommand{\Borxi}{{\it\Delta}^{0}_{\xi}}
\newcommand{\Boraxi}{{\it\Sigma}^{0}_{\xi}}
\newcommand{\Bormxi}{{\it\Pi}^{0}_{\xi}}

\newtheorem{thm} {Theorem}
\newtheorem{defi} [thm] {Definition}
\newtheorem{cor} [thm] {Corollary}
\newtheorem{lem} [thm] {Lemma}
\newtheorem{prop} [thm] {Proposition}

\begin{document}

\maketitle

\centerline{$\bullet$ CNRS, Universit\'e de Paris, Sorbonne Universit\'e,} 

\centerline{Institut de Math\'ematiques de Jussieu-Paris Rive Gauche, Equipe de Logique Math\'ematique}

\centerline{Campus des Grands Moulins, b\^atiment Sophie-Germain, case 7012, 75205 Paris cedex 13, France}

\centerline{finkel@math.univ-paris-diderot.fr}\medskip

\centerline{$\bullet^1$ Sorbonne Universit\'e, Universit\'e de Paris, CNRS,} 

\centerline{Institut de Math\'ematiques de Jussieu-Paris Rive Gauche, Equipe d'Analyse Fonctionnelle}

\centerline{Campus Pierre et Marie Curie, case 247, 4, place Jussieu, 75 252 Paris cedex 5, France}

\centerline{dominique.lecomte@upmc.fr}\medskip

\centerline{$\bullet^1$ Universit\'e de Picardie, I.U.T. de l'Oise, site de Creil,}

\centerline{13, all\'ee de la fa\"\i encerie, 60 107 Creil, France}\medskip\medskip\medskip\medskip\medskip\medskip


\ufootnote{{\it Keywords and phrases.}~Languages of finite or infinite words, context-free, one-counter automaton, $\omega$-power, topological complexity, Borel class, complete set}

\noindent {\bf Abstract.} The $\omega$-power  of a  finitary language $L$ over a finite alphabet $\Si$  is the language of infinite words over $\Si$ defined by
$$L^\infty\! :=\!\{ w_0w_1\ldots\!\in\!\Si^\om\mid\fa i\!\in\!\omega  ~~w_i\!\in\! L\} .$$ 
The $\om$-powers appear very naturally in Theoretical Computer Science in the characterization of several classes of languages of infinite words accepted by various kinds of automata, like 
 B\"uchi automata or  B\"uchi pushdown automata. 
 We survey some recent results about the links relating Descriptive Set Theory and $\om$-powers.

\newpage

\section{$\!\!\!\!\!\!$ Introduction}\indent

In the sixties, B\"uchi  studied acceptance of  infinite words by finite automata with the now called  B\"uchi  acceptance condition, in order 
to prove the decidability of the monadic second order theory of one successor
over the integers.  Since then there has been a lot of work on  regular $\omega$-languages, accepted by   B\"uchi automata, or by some other variants of automata over infinite words, like Muller or Rabin automata, and by other  finite machines, like 
pushdown automata, counter automata, Petri nets, Turing machines, \ldots, with various acceptance conditions, 
see \cite{Thomas90,Staiger97,PerrinPin}.\medskip

 The class of regular $\om$-languages, those acccepted by B\"uchi automata, is the $\om$-Kleene closure 
of the family $REG$ of regular finitary languages.  The $\om$-{\bf Kleene closure} of a class of languages of finite words over finite alphabets is the class of $\om$-languages of the form $\bigcup_{1\leq j \leq n}~U_j\cdot V_j^\infty$, for some regular  finitary languages $U_j$ and $V_j$, $1\leq j \leq n$, where for any finitary  language $L\!\subseteq\!\Sigma^{<\om}$ over the alphabet $\Si$, the $\om$-{\bf power} $L^\infty$ of $L$ is  the set of the infinite words constructible with $L$ by concatenation, i.e., 
$$L^\infty\! :=\!\{\ w_0w_1\ldots\!\in\!\Sigma^\omega\mid\forall i\!\in\!\omega~\ w_i\!\in\! L\ \} .$$
Note that we denote here $L^\infty$ the $\omega$-power associated with $L$, as in \cite{Lecomte05,Fink-Lec2}, while it is often denoted $L^\om$ in Theoretical Computer Science papers, as in \cite{Staiger97,Fin01a,Fin03a,Fin-Lec}. Here we reserved the notation $L^\om$ to denote the Cartesian product of countably many copies of $L$ since this will be often used in this paper.\medskip
      
  Similarly,  the operation of taking the $\om$-power of a finitary language appears in the characterization of the class of context-free\ol s 
as the $\om$-Kleene closure of the family  of context-free finitary languages (we refer the reader to \cite{ABB96} for basic notions about context-free languages). And the class of  $\om$-languages accepted by B\"uchi one-counter automata  is also  the $\om$-Kleene closure of the family of  finitary languages accepted by one-counter automata. Therefore the operation $L\!\ra\! L^\infty$  is a fundamental operation over finitary languages leading to \ol s. The $\om$-powers of regular languages have been studied in \cite{LT87,Staiger97}.\medskip 

  During the last years, the $\om$-powers have  been studied from the perspective of Descriptive Set Theory 
in a few papers \cite{Fin01a,Fin03a,Fin04-FI,Lecomte05,Fin-Dup06,Fin-Lec,Fink-Lec2,Fink-Lec3}.  We mainly review these recent works in the present survey. \medskip

 Since the set $\Sio$ of infinite words over a finite alphabet $\Si$ can be   equipped 
with the usual Cantor topology, the question of  the topological  complexity of  $\om$-powers of 
finitary  languages, from the point of view of descriptive set theory,  naturally arises and has  been posed by 
Niwinski \cite{Niwinski90},  Simonnet \cite{Simonnet92},  and  Staiger \cite{Staiger97}.\medskip

 As the concatenation map, from $L^\omega$ onto $L^\infty$, which associates to a given sequence $(w_i)_{i\in\omega}$ of finite words the concatenated word $w_0w_1\ldots$, is continuous, an $\omega$-power is always an analytic set. It was proved in \cite{Fin03a} that there exists a (context-free) language $L$ such that $L^\infty$ is analytic but not Borel. Amazingly, the language $L$ is  very simple to describe and it is accepted by a simple one-counter automaton. Louveau has proved independently that analytic-complete $\omega$-powers exist, but the existence was proved in a non effective way (this is non-published work).\medskip

 One of our first tasks was to study the position of $\om$-powers with respect to the Borel hierarchy (and beyond to the projective hierarchy).  A characterization of $\om$-powers in the  Borel classes ${\bf \Si}^0_1$, ${\bf \Pi}^0_1$  and  ${\bf \Pi}^0_2$ has been given by Staiger in \cite{Staiger97-2}.\medskip

 Concerning Borel $\om$-powers, it was proved that, for each integer $n\geq 1$, there exist some $\om$-powers of  (context-free) languages which are ${\bf \Pi}_n^0$-complete Borel sets,  \cite{Fin01a}.   It was proved in \cite{Fin04-FI} that  there exists a finitary language $L$ such that $L^\infty$ is a Borel set of infinite rank, and in \cite{Fin-Dup06} that there is a (context-free) language $W$ such that $W^\infty$ is Borel above ${\bf\Delta}_\omega^0$. We recently proved that there are complete $\om$-powers of one-counter languages, for every Borel class of finite rank, \cite{Fink-Lec3}. \medskip

 We  proved  in \cite{Fin-Lec, Fink-Lec2}  a result which showed that $\omega$-powers exhibit a great topological complexity: 
for each countable ordinal $\xi\!\geq\! 1$, there are $\bormxi$-complete $\omega$-powers, and $\boraxi$-complete 
$\omega$-powers.  This result has an effective aspect: for each recursive ordinal $\xi\! <\!\omega_1^{\text{CK}}$, where 
$\omega_1^{\text{CK}}$ is the first non-recursive ordinal, there are recursive finitary languages $P$ and $S$ such that $P^\infty$ is 
$\bormxi$-complete and $S^\infty$ is $\boraxi$-complete.\medskip

 Many questions are still open about the topological complexity of $\om$-powers of languages in a given class like the class of context-free languages, one-counter languages, recursive languages, or more generally  languages accepted by some kind of automata over finite words. We mention some of these open questions in this paper. \medskip

 This article is organized as follows.  Some basic notions of topology are recalled in Section 2. Notions of automata and formal language theory are recalled in Section 3, and $\om$-powers of finitary languages accepted by automata are studied in this section. The study of 
$\om$-powers of finitary languages in the classical setting of descriptive set theory forms Section 4. Finally, we provide in Section 5 some complexity results about some sets of finitary languages whose associated $\om$-power is in some class of sets.

\section{$\!\!\!\!\!\!$ Topology}\indent

 When $\Si$ is a finite alphabet, a nonempty {\bf finite word} over $\Si$ is a sequence $w\! =\! a_0\ldots a_{l-1}$, where $a_i\!\in\!\Sigma$ for each $i\! <\! l$, and $l\geq 1$ is a natural number. The {\bf length} of $w$ is $l$, denoted by $|w|$. A word of length one is of the form $(a)$. The {\bf empty word} is denoted by $\lambda$ and satisfies $|\lambda |\! =\! 0$. When $w$ is a finite word over $\Sigma$, we write 
$w\! =\! w(0)w(1)\ldots w(l\! -\! 1)$, and the prefix $w(0)w(1)\ldots w(i\! -\! 1)$ of $w$ of length $i$ is denoted by $w\vert i$, for any 
$i\!\leq\! l$. We also write $u\!\subseteq\! v$ when the word $u$ is a prefix of the finite word $v$. The set of finite words over $\Si$ is denoted by $\Si^{<\om}$, and $\Si^+$ is the set of nonempty finite words over $\Sigma$. A (finitary) {\bf language} over $\Sigma$ is a subset of $\Sis$. For $L\!\subseteq\!\Sis$, the {\bf complement} $\Sis\!\setminus\! L$ of $L$ (in $\Sis$) is denoted by $L^-$. We sometimes write $a$ for $\{ (a)\}$, for short.\medskip

  The first infinite ordinal is $\om$. An $\om$-{\bf word} over $\Si$ is an $\om$-sequence $a_0a_1\ldots$, where 
$a_i\!\in\!\Sigma$ for each natural number $i$. When $\sigma$ is an $\om$-word over $\Si$, the length of $\sigma$ is 
$\vert\sigma\vert\! =\!\om$, and we write $\sigma =\sigma(0)\sigma(1)\ldots$, and the prefix 
$\sigma(0)\sigma(1)\ldots\sigma(i\! -\! 1)$ of $\sigma$ of length $i$ is denoted by $\sigma\vert i$, for any natural number $i$. We also write $u\!\subseteq\!\sigma$ when the finite word $u$ is a prefix of the $\om$-word $\sigma$. The set of $\om$-words over $\Si$ is denoted by $\Si^\om$. An $\om$-{\bf language} over $\Sigma$ is a subset of  $\Si^\om$. For 
$A\subseteq \Si^\om$, the complement $\Si^\om\!\setminus\! A$ of $A$ is denoted by $A^-$.\medskip

 The usual {\bf concatenation} product of two finite words $u$ and $v$ is denoted $u^\frown v$ (and sometimes just $uv$). This product is extended to the product of a finite word $u$ and an $\om$-word $\sigma$: the infinite word $u^\frown\sigma$ is then the $\om$-word such that $(u^\frown\sigma )(k)\! =\! u(k)$ if $k\! <\! |u|$, and $(u^\frown\sigma )(k)\! =\!\sigma (k\! -\! |u|)$ if $k\!\geq\! |u|$.\medskip

 If $E$ is a set, $l\!\in\!\omega$ and $(e_i)_{i<l}\!\in\! E^l$, then ${^\frown}_{i<l}\ e_i$ is the concatenation 
$e_0\ldots e_{l-1}$. Similarly, ${^\frown}_{i\in\omega}\ e_i$ is the concatenation $e_0e_1\ldots$ For $L\subseteq\Sis$, 
$L^\infty\! :=\!\{ \sigma =w_0w_1\ldots\!\in\!\Si^\om\mid\fa i\!\in\!\omega  ~~w_i\!\in\! L\}$ is the $\om$-{\bf power} of $L$.\medskip

 We now recall some notions of topology, assuming  the reader to be familiar with the basic notions, that can be found in  \cite{Moschovakis80,Kechris94,Staiger97,PerrinPin}. The topological spaces in which we will work in this paper will be subspaces of 
$\Sigma^\omega$, where $\Si$ is either finite having at least two elements (like $2\! :=\!\{ {\bf 0},{\bf 1}\}$), or countably infinite. Note that here $2$ is considered as an alphabet, and we will do it also for 3,4; sometimes, we will view it as a letter, and in this case we will denote it by {\bf 2}, like we just did it for {\bf 0},{\bf 1}. The topology on $\Sigma^\omega$ is the product topology of the discrete topology on 
$\Sigma$. For $w\!\in\!\Si^{<\om}$, the set defined by $N_w\! :=\!\{\alpha\!\in\!\Sigma^\omega\mid w\!\subseteq\!\alpha\}$ is a basic clopen (i.e., closed and open) set of $\Sigma^\omega$. The open subsets of $\Sio$ are of the form 
$W^\frown\Si^\om\! :=\!\{ w\sigma\mid w\!\in\! W\mbox{ and }\sigma\!\in\!\Si^\om\}$, where $W\!\subseteq\!\Si^{<\om}$. When $\Si$ is finite, this topology is called the {\bf Cantor topology} and $\Sio$ is compact. When $\Si\! =\!\omega$, $\Sigma^\omega$ is the Baire space, which is homeomorphic to 
$\mathbb{P}_\infty\! :=\!\{\alpha\!\in\! 2^\omega\mid\forall i\!\in\!\omega\ \exists j\!\geq\! i\ \ \alpha (j)\! =\! {\bf 1}\}$, via the map defined on 
$\omega^\omega$ by $h(\beta )\! :=\! {\bf 0}^{\beta (0)}{\bf 1}{\bf 0}^{\beta (1)}{\bf 1}\ldots$ There is a natural metric on $\Sio$, the 
{\bf prefix metric} defined as follows. For $\sigma\!\not=\!\tau\!\in\!\Sio$, $d(\sigma ,\tau )\! :=\! 2^{-l_{pref(\sigma ,\tau )}}$, where $l_{pref(\sigma ,\tau )}$ is the first natural number $n$ such that $\sigma (n)\!\not=\!\tau (n)$. The topology induced on $\Sigma^\omega$ by this metric is our topology.\medskip

 We now define the Borel hierarchy.

\begin{defi}
Let $X$ be a topological space, and $n\!\geq\! 1$ be a natural number. The classes ${\bf\Si}_n^0(X)$ and ${\bf\Pi}_n^0(X) $ of the {\bf Borel hierarchy} are inductively defined as follows:\smallskip

${\bf \Si}^0_1(X) $ is the class of open subsets of $X$.\smallskip

${\bf \Pi}^0_1(X) $ is the class of closed subsets of $X$.\smallskip

${\bf \Si}^0_{n+1}(X)$ is the class of countable unions of ${\bf \Pi}^0_n$-subsets of  $X$.\smallskip

${\bf \Pi}^0_{n+1}(X)$ is the class of countable intersections of ${\bf \Si}^0_n$-subsets of $X$.\smallskip

\noindent The Borel hierarchy is also defined for the transfinite levels. Let $\xi\!\geq\! 2$ be a countable ordinal.\smallskip

${\bf \Si}^0_\xi (X)$ is the class of countable unions of subsets of $X$ in $\bigcup_{\gamma <\xi}~{\bf \Pi}^0_\gamma$.\smallskip

${\bf \Pi}^0_\xi (X)$ is the class of countable intersections of subsets of $X$ in $\bigcup_{\gamma <\xi}~{\bf \Si}^0_\gamma$.
\end{defi}

 Suppose now that $\xi\!\geq\! 1$ is a countable ordinal and $X\!\subseteq\! Y$, where $X$ is equipped with the induced topology. Then 
$\boraxi (X)\! =\!\{ A\cap X\mid A\!\in\!\boraxi (Y)\}$, and similarly for $\bormxi$, see \cite[Section 22.A]{Kechris94}. Note that we defined the Borel classes ${\bf\Si}^0_\xi(X)$ and ${\bf\Pi}^0_\xi (X)$ mentioning the space $X$. However, when the context is clear, we will sometimes omit $X$ and denote ${\bf \Si}^0_\xi(X)$ by ${\bf \Si}^0_\xi$ and similarly for the dual class. The Borel classes are closed under finite intersections and unions, and continuous preimages. Moreover, $\boraxi$ is closed under countable unions, and $\bormxi$ under countable intersections. As usual, the ambiguous class $\borxi$ is the class $\boraxi\cap\bormxi$. The class of {\bf Borel sets} is 
$$\borel\! :=\!\bigcup_{1\leq\xi <\omega_1}\ \boraxi\! =\!\bigcup_{1\leq\xi <\omega_1}\ \bormxi\mbox{,}$$ 
where $\om_1$ is the first uncountable ordinal. The {\bf Borel hierarchy} is as follows:
$$\begin{array}{ll}  
& \ \ \ \ \ \ \ \ \ \ \ \ \ \ \ \ \ \ \ \ \ \ \ \ \ \boraone\! =\!\hbox{\rm open}\ \ \ \ \ \ \ \ \ \ \ \ \ 
\boratwo\! \ \ \ \ \ \ \ \  \ \ \ 
\ldots\ \ \ \ \ \ \ \ \ \ \ \ \boraom\ \ \ \ \ \ldots\cr  
& \borone\! =\!\hbox{\rm clopen}\ \ \ \ \ \ \ \ \ \ \ \ \ \ \ \ \ \ \ \ \ \ \ \ \ \ \ 
\bortwo\ \ \ \ \ \ \ \ \ \ \ \ \ \ \ \ \ \ \ \ \ \ \ \ \ \ \ \ \ \ \ \borom\ \ \ \ \ \ \ \ \ \ \ \ \ \ \ \ \ \ \ \ \ \ \borel\cr
& \ \ \ \ \ \ \ \ \ \ \ \ \ \ \ \ \ \ \ \ \ \ \ \ \ \bormone\! =\!\hbox{\rm closed}\ \ \ \ \ \ \ \ \ \ \bormtwo\! \ \ \ \  \ \ \ \ \ \ \ \ \ldots
\ \ \ \ \ \ \ \ \ \ \ \ \bormom\ \ \ \ \ \ldots
\end{array}$$

 This picture means that any class is contained in every class at the right of it, and the inclusion is strict in any of the spaces 
$\Sigma^\omega$. A subset of $\Si^\om$ is a Borel set of {\bf rank} $\xi$ if it is in ${\bf\Si}^0_\xi\cup {\bf\Pi}^0_\xi$ but not in 
$\bigcup_{1\leq\gamma <\xi}~({\bf\Si}^0_\gamma\cup {\bf\Pi}^0_\gamma)$.\medskip

 We now define completeness with respect to reducibility by continuous functions. Let $Y,\Si$ be finite alphabets, $A\!\subseteq\! Y^\om$ and $C\!\subseteq\! \Si^\om$. We say that $A$ is {\bf Wadge reducible} to $C$ if there exists a continuous function 
$f\! :\! Y^\om\!\ra\!\Si^\om$ such that $A\! =\! f^{-1}(C)$. Now let ${\bf\Gamma}$ be a class of sets closed under continuous pre-images like $\boraxi$ or $\bormxi$. A subset $C$ of $\Si^\om$ is said to be ${\bf\Gamma}$-{\bf hard} if, for any finite alphabet $Y$ and any 
$A\!\subseteq\! Y^\om$, $A\!\in\! {\bf\Gamma}$ implies that  $A$ is Wadge reducible to $C$. If moreover $C$ is in ${\bf\Gamma}(\Si^\om )$, then we say that $C$ is ${\bf\Gamma}$-{\bf complete}. The ${\bf\Si}^0_n$-complete sets and the ${\bf\Pi}^0_n$-complete sets are thoroughly characterized in \cite{Staiger86a}. Recall that a subset of $\Sio$ is ${\bf\Si}^0_\xi$ (respectively ${\bf\Pi}^0_\xi$)-complete if and only if it is in ${\bf\Si}^0_\xi$ but not in ${\bf\Pi^0_\xi}$ (respectively in ${\bf\Pi}^0_\xi$ but not in ${\bf\Si}^0_\xi$), and that such sets exist (see \cite{Kechris94}). For example, the singletons of $2^\omega$ are $\bormone$-complete. The set $\mathbb{P}_\infty$ defined at the beginning of the present section is a well known example of a $\bormtwo$-complete set. We say that $\bf\Gamma$ is a 
{\bf Wadge class} if there is a $\bf\Gamma$-complete set. The {\bf Wadge hierarchy} of Borel sets given by the inclusion of these classes is a great refinement of the Borel hierarchy of the classes $\boraxi$ and $\bormxi$. Among the new classes appearing in this hierarchy, we can mention the classes of transfinite differences of $\boraxi$ sets. If $\eta$ is a countable ordinal and $(A_\theta )_{\theta <\eta}$ is an increasing sequence of subsets of some set $X$, then we set 
$$D_\eta\big( (A_\theta )_{\theta <\eta} \big) \! :=\!\{ x\!\in\! X\mid\exists\theta\! <\!\eta~\ \ 
x\!\in\! A_\theta\!\setminus\!\bigcup_{\theta'<\theta}\ A_{\theta'}\ \ \hbox{\rm and\ the\ parity\ of}\ \ \theta\ \ \hbox{\rm is opposite\ to\ that\ of}\ \ \eta\}.$$
If moreover $\xi\!\geq\! 1$ is a countable ordinal, then we set 
$D_\eta (\boraxi )\! :=\!\big\{ D_\eta\big( (A_\theta )_{\theta <\eta}\big)\mid  \forall  \theta <\eta ~~ A_\theta\!\in\!\boraxi\big\}$.\medskip

 The class $\check {\bf\Gamma}\! :=\!\{\neg A\mid A\!\in\! {\bf\Gamma}\}$ is the class of the complements of the sets in ${\bf\Gamma}$, and is called the {\bf dual class} of $\bf\Gamma$. In particular, $\check {{\bf\Si}^0_\xi}\! =\! {\bf\Pi}^0_\xi$ and 
$\check {{\bf\Pi}^0_\xi}\!=\! {\bf\Si}^0_\xi$.\medskip

 There are some subsets of the topological space $\Sio$ which are not Borel sets. In particular, there is another hierarchy beyond the Borel hierarchy,  called the projective hierarchy. The first class of the projective hierarchy is the class ${\bf\Si}^1_1$ of analytic sets. A subset $A$ of $\Sio$ is {\bf analytic} if we can find a finite alphabet $Y$ and a Borel subset $B$ of 
$(\Si\!\times\! Y)^\om$ such that $x\!\in\! A\Leftrightarrow\exists y\!\in\! Y^\om ~(x, y)\!\in\! B$, where 
$(x, y)\!\in\! (\Si\!\times\! Y)^\om$ means that $(x, y)(i)\! =\!\big( x(i),y(i)\big)$ for each natural number $i$. A subset of $\Sigma^\omega$ is analytic if it is empty, or the image of the Baire space by a continuous map. The class ${\bf\Si}^1_1$ of analytic sets contains the class of Borel sets in any of the spaces $\Sigma^\omega$. Note that ${\bf\Delta}_1^1\! =\! {\bf\Si}^1_1 \cap {\bf\Pi}^1_1$, where 
${\bf\Pi}^1_1\! :=\!\check\ana$ is the class of {\bf co-analytic} sets, i.e., of complements of analytic sets. Similarly, the class of projections of $\ca$ sets is denoted ${\bf\Sigma}^1_2$.\medskip
 
 The $\om$-power of a finitary language $L$ is always an analytic set. Indeed, if $L$ is finite and has $n$ elements, then $L^\infty$  is the continuous image of the compact set $\{ {\bf 0},{\bf 1},\ldots ,{\bf n\! -\! 1}\}^\om$. If $L$ is infinite, then there is a bijection between 
$L$ and $\om$, and $L^\infty$ is the continuous image of the Baire space $\om^\om$, \cite{Simonnet92}.

\section{$\!\!\!\!\!\!$ Complexity of $\om$-powers of languages accepted by automata}

\subsection{$\!\!\!\!\!\!$ Automata}\label{section-automata}\indent

 We assume the reader to be familiar with formal \la s, see for example \cite{HopcroftMotwaniUllman2001,Thomas90}.

\vfill\eject 
 
 We first recall some of the definitions and results concerning automata, pushdown automata, regular and context-free  \la s, as presented in \cite{ABB96,cg,Staiger97}.

\begin{defi} 
A {\bf pushdown automaton} is a 7-tuple $\mathcal{A}\! =\! (Q,\Si ,\Ga ,q_0,Z_0,\delta ,F)$, where $Q$ is a finite set of states, 
$\Sigma$ is a finite input alphabet, $\Gamma$ is a finite pushdown alphabet, $q_0\!\in\! Q$ is the initial state, $Z_0\!\in\!\Ga$ is the start symbol which is the bottom symbol and always remains at the bottom of the pushdown stack, $\delta$ is a map from 
$Q\!\times\! (\Si\cup\{\lambda\} )\!\times\!\Ga$ into the set of finite subsets of $Q\!\times\!\Gas$, and $F\!\subseteq\! Q$ is the set of final states. The automaton $\mathcal{A}$ is said to be {\bf deterministic} if  $\delta$ is a map from 
$Q\!\times\! (\Si\cup\{\lambda\} )\!\times\!\Ga$ into the set of  subsets of cardinal one, i.e., singletons, of $Q\!\times\!\Gas$. The automaton $\mathcal{A}$ is said to be {\bf real-time} if there is no $\lambda$-transition, i.e., if $\delta$ is a map from 
$Q\!\times\!\Si\!\times\!\Ga$ into the set of finite subsets of $Q\!\times\!\Gas$.\smallskip

 If $\gamma\in\Ga^+$ describes the pushdown stack content, then the leftmost symbol will be assumed to be on the ``top" of the stack. A {\bf configuration} of the pushdown automaton $\mathcal{A}$ is a pair $(q,\gamma )$, where $q\!\in\! Q$ and 
$\gamma\in\Gas$. For $a\!\in\!\Si\cup\{\lambda\}$, $\gamma ,\beta\!\in\!\Gas$ and $Z\!\in\!\Ga$, if $(p,\beta )$ is in $\delta (q,a,Z)$, then we write $a\! :\! (q,Z\gamma )\!\mapsto_{\mathcal{A}}\! (p,\beta\gamma )$.\smallskip
  
 Let $w\! =\! a_0\ldots a_{l-1}$ be a  finite word over $\Si$. A  sequence of configurations $r\! =\! (q_i,\gamma_i)_{i<N}$ is called a  {\bf run of $\mathcal{A}$ on $w$ starting in the configuration} $(p,\gamma )$ if
\begin{enumerate}
\ite[(1)] $(q_0,\gamma_0)\! =\! (p,\gamma)$,

\ite[(2)]  for each $i\! <\! N\! -\! 1$, there exists $b_i\in\Si\cup\{\lambda\}$ satisfying 
$b_i\! :\! (q_i,\gamma_i)\!\mapsto_{\mathcal{A}}\! (q_{i+1},\gamma_{i+1})$ such that $a_0\ldots a_{l-1}\! =\! b_0\ldots b_{N-2}$.
\end{enumerate}

 A run $r$ of $\mathcal{A}$ on $w$ starting in configuration $(q_0,Z_0)$ will be simply called a {\bf run of} $\mathcal{A}$ {\bf on} 
$w$. The run is {\bf accepting} if it ends in a final state.\smallskip

 The language $L(\mathcal{A})$ {\bf accepted} by $\mathcal{A}$ is the set of words admitting an accepting run by $\mathcal{A}$. A 
{\bf context-free language} is a finitary language which is accepted by a pushdown automaton. We denote by $CFL$ the class of context-free languages.\smallskip

 If we omit the pushdown stack in the definition of a pushdown automaton, we get the notion of a (finite state) automaton. Note that every finite state automaton is equivalent to a deterministic real-time finite state automaton. A {\bf regular  language} is a finitary language which is accepted by a (finite state)   automaton. We denote by $REG$ the class of regular  languages.\smallskip

 A {\bf one-counter automaton} is a pushdown automaton with a pushdown alphabet of the form 
${\Ga\! =\! \{ Z_0,z\}}$, where $Z_0$ is the bottom symbol and always remains at the bottom of the pushdown stack. A 
{\bf one-counter language} is a (finitary) language which is accepted by a one-counter automaton. 
\end{defi}

\begin{defi} Let $\Sigma ,\Gamma$ be finite alphabets.\smallskip

(a) A $(\Sigma ,\Gamma )$-{\bf substitution} is a map $f\! :\!\Sigma\!\rightarrow\! 2^{\Gamma^{<\omega}}$.\smallskip

(b) We extend this map to $\Sigma^{<\omega}$ be setting 
$f({^\frown}_{i<l}\ a_i)\! :=\!\{ {^\frown}_{i<l}\ w_i\mid\forall i\! <\! l~~w_i\!\in\! f(a_i)\}$, where 
$l\!\in\!\omega$ and $a_0,\cdots ,a_{l-1}\!\in\!\Sigma$.\smallskip

(c) We further extend this map to $2^{\Sigma^{<\omega}}$ by setting $f(L)\! :=\!\bigcup_{w\in L}~f(w)$.\smallskip

(d) Let $f$ be a $(\Sigma ,\Gamma )$-substitution, and $\mathcal{F}$ be a family of languages. If the language $f(a)$ belongs to 
$\mathcal{F}$ for each $a\!\in\!\Si$, then the substitution $f$ is called a $\mathcal{F}$-{\bf substitution}.\smallskip

(e) We then define the operation $\square$ on families of languages. Let $\mathcal{E}$, $\mathcal{F}$ be families of (finitary) languages. Then 
$\mathcal{E}~\square ~\mathcal{F}\! :=\!\{ f(L)\mid L\!\in\!\mathcal{E}\mbox{ and }f\mbox{ is a }\mathcal{F}\mbox{-substitution}\}$. 
\end{defi}

  The operation of substitution gives rise to an infinite hierarchy of context-free finitary languages defined as follows.

\begin{defi} Let  $OCL(0)=REG$ be the class of regular languages,   $OCL(1)=OCL$  be  the class of one-counter languages, 
and $OCL(k\! +\! 1)\! =\! OCL(k)~\square ~OCL$, for $k\!\geq\! 1$.\end{defi}

 It is well known that the hierarchy given by the families of languages $OCL(k)$ is strictly increasing. And there is a characterization of these languages in terms of automata.

\begin{prop}[\cite{ABB96}]  A language $L$ is in $OCL(k)$ if and only if $L$ is recognized by a pushdown automaton such that, during any computation, the words in the pushdown stack remain in a language of the form 
$(z_{k-1})^{<\om}\ldots (z_0)^{<\om}Z_0$, where $\{ Z_0, z_0,\ldots ,z_{k-1}\}$ is the pushdown alphabet. Such an automaton is called a $k$-{\bf iterated counter automaton}. The union $ICL\! :=\!\bigcup_{k\geq 1}~OCL(k)$ is called the family of {\bf iterated counter languages}, which is the closure under substitution of the family $OCL$.\end{prop}

\subsection{$\!\!\!\!\!\!$  ${\bf \Pi}_n^0$-complete and ${\bf \Si}_n^0$-complete $\om$-powers}\label{Si0N-Pi0N}\indent

 Wadge first gave a description of the Wadge hierarchy of Borel sets, see \cite{Wadge83}. Duparc got in \cite{Duparc01} a new proof of Wadge's results in the case of  Borel sets of finite rank, and he gave a normal form of  Borel sets of finite rank, i.e., an inductive construction of a Borel set of every given degree. His proof relies on set theoretic operations which are the counterpart of arithmetical operations over ordinals needed to compute the Wadge degrees.\medskip
 
 In fact J. Duparc studied the Wadge hierarchy via the study of the conciliating hierarchy. He introduced in \cite{Duparc01} the conciliating sets, which are sets of finite {\it or} infinite words over an alphabet $\Si$, i.e., subsets of 
$\Si^{\leq\om}\! :=\!\Si^{<\om}\cup\Si^\om$. In particular, the  set theoretic operation  of exponentiation, defined over concilating sets, 
has been very useful in the study of context-free $\omega$-powers.\medskip 
 
 We first recall the following.

\begin{defi}
Let  $\Si_A$ be a finite alphabet, $\tla$ be a letter out of $\Si_A$, $\Si\! :=\!\Si_A\cup\{\tla\}$, and $x$ be a finite or infinite word over the alphabet $\Si$. Then $x^\tla$ is inductively defined as follows.\smallskip

- $\lambda^\tla\! :=\!\lambda$.\smallskip

- For a finite word $u\!\in\!\Si^{<\om}$, 
$\left\{\!\!\!\!\!\!\!
\begin{array}{ll}
& (ua)^\tla\! :=\! u^\tla a\mbox{ if }a\!\in\!\Si_A\mbox{,}\cr
& (u\tla)^\tla\! :=\! u^\tla\mbox{ with its last letter removed if }|u^\tla|\! >\! 0\mbox{,}\cr
& (u\tla)^\tla\! :=\!\lambda\mbox{ if }|u^\tla|\! =\! 0.
\end{array}
\right.$\smallskip

- For an infinite word $\sigma$, $\sigma^\tla\! :=\!\mbox{lim}_{n\in\om}~(\sigma\vert n)^\tla$, where, given 
$(w_n)\!\in\! (\Si_A^{<\om})^\om$ and $w\!\in\!\Si_A^{<\om}$, 
$$w\!\subseteq\!\mbox{lim}_{n\in\om}~w_n\Leftrightarrow\exists p\!\in\!\omega ~~\fa n\!\geq\! p~~w_n\vert |w|\! =\! w.$$
\end{defi}

\noindent\bf Remark.\rm\ For $x\!\in\!\Si^{\leq\om}$, $x^\tla$ denotes the string $x$, once every $^\tla$ occuring in $x$ has been ``evaluated" as the back space operation (the one familiar to your computer!), proceeding from left to right inside $x$. In other words, $x^\tla\! =\! x$ from which every interval of the form $``a\tla "$ ($a\!\in\!\Si_A$) is removed.\medskip

 For example, if $x\! =\! (a\tla)^n$ for some $n\!\geq\! 1$, $x\! =\! (a\tla)^\om$ or $x\! =\! (a\tla\tla)^\om$ then $x^\tla\! =\!\lambda$. If  $x\! =\! (ab\tla)^\om$, then $x^\tla\! =\! a^\om$. If $x\! =\! bb(\tla a)^\om$, then $x^\tla\! =\! b$.\medskip 

 We now can define the operation $A\!\mapsto\! A^\sim$ of exponentiation of conciliating sets.

\begin{defi}
Let  $\Si_A$ be a finite alphabet, $\tla$ be a letter out of $\Si_A$, $\Si\! :=\!\Si_A\cup\{\tla\}$, and $A\!\subseteq\!\Si_A^{\leq\om}$. Then we set $A^\sim\! :=\!\{ x\!\in\!\Si^{\leq\om}\mid x^\tla\!\in\! A\}$.
\end{defi}

 Roughly speaking, the operation $\sim$ is monotone with regard to the Wadge ordering and produces some sets of higher complexity.\medskip 
 
The first author proved in \cite{Fin01a} that the class $CFL_\om$ of context-free $\om$-languages, (i.e., those which are accepted by   pushdown automata with a B\"uchi acceptance condition expressing that ``some final state appears infinitely often during an infinite computation''), is closed under this operation $\sim$.\medskip 
  
 We now recall a  slightly modified variant of the operation $\sim$, introduced in \cite{Fin01a}, and which is particularly suitable to infer properties of $\om$-powers. 
  
\begin{defi}\label{approx}
Let  $\Si_A$ be a finite alphabet, $\tla$ be a letter out of $\Si_A$, $\Si\! :=\!\Si_A\cup\{\tla\}$, and $A\!\subseteq\!\Si_A^{\leq\om}$. Then we set $A^\approx\! :=\!\{ x\!\in\!\Si^{\leq\om}\mid x^\tla\!\in\! A\}$, where $x^\tla$ is inductively defined as follows.\smallskip

- $\lambda^\tla\! :=\!\lambda$.\smallskip

- For a finite word $u\!\in\!\Si^{<\om}$, 
$\left\{\!\!\!\!\!\!\!
\begin{array}{ll}
& (ua)^\tla\! :=\! u^\tla a\mbox{ if }a\!\in\!\Si_A\mbox{,}\cr
& (u\tla)^\tla\! :=\! u^\tla\mbox{ with its last letter removed if }|u^\tla|\! >\! 0\mbox{,}\cr
& (u\tla)^\tla\mbox{ is undefined if }|u^\tla|\! =\! 0.
\end{array}
\right.$\smallskip

- For an infinite word $\sigma$, $\sigma^\tla\! :=\!\mbox{lim}_{n\in\om}~(\sigma\vert n)^\tla$.
\end{defi}

 The only difference is that here $(u\tla)^\tla$ is undefined if $|u^\tla|\! =\! 0$. It is easy to see that if $A\!\subseteq\!\Si_A^\om$ is a Borel set such that $A\!\neq\!\Si_A^\om$, i.e., $A^-\!\neq\!\emptyset$, 
then $A^\approx$ is Wadge equivalent to $A^\sim$ (see \cite{Fin01a}) and this implies the following result: 

\begin{thm}\label{thedup2}
Let  $\Si_A$ be a finite alphabet, and $n\!\geq\! 2$ be a natural number. If $A\!\subseteq\!\Si_A^\om$ is ${\bf \Pi}_n^0$-complete, 
then $A^\approx$ is ${\bf \Pi}_{n+1}^0$-complete.
\end{thm}

\noi\bf Notation.\rm\ Let $\Si_A$ be a finite alphabet, $\tla$ be a letter out of $\Si_A$, and $\Si\! :=\!\Si_A\cup\{\tla\}$. The language $L_3$ over $\Si$ is the context-free language generated by the context-free grammar with the following production rules:
$$\begin{array}{ll}
& S\!\ra\! aS\tla S\mbox{ with }a\!\in\!\Si_A\mbox{,}\cr
& S\!\ra\! a\tla S\mbox{ with }a\!\in\!\Si_A\mbox{,}\cr
& S\!\ra\!\lambda 
\end{array}$$
(see \cite{HopcroftMotwaniUllman2001} for the basic notions about grammars). This language $L_3$ corresponds to the words where every letter of $\Si_A$ has been removed after using the backspace operation. It is easy to see that $L_3$ is a \de one-counter \la , i.e., $L_3$ is accepted by a \de one-counter automaton. Moreover,  for $a\!\in\!\Si_A$, the language $L_3a$ is also accepted by a deterministic one-counter automaton.\medskip

 We can now state the following result, which implies that the class of $\om$-powers is closed under the operation  $A \ra A^\approx$. 
 
\begin{lem}[see \cite{Fin01a}] Whenever $A\!\subseteq\!\Si_A^\om$, the \ol~ $A^\approx\!\subseteq\!\Si^\om$ is obtained by substituting in $A$ the language $L_3a$ for each letter $a\!\in\!\Si_A$.\end{lem} 
  
   An $\om$-word $\sigma\!\in\! A^\approx$ may be considered as an $\om$-word $\sigma^\tla\!\in\! A$ to which we possibly add, before the first letter $\sigma^\tla (0)$ of $\sigma^\tla$ (respectively, between two consecutive letters $\sigma^\tla (n)$ and 
$\sigma^\tla (n\! +\! 1)$ of $\sigma^\tla$), a finite word belonging to the context-free (finitary) language $L_3$. 
  
\begin{cor} \label{presop}
Whenever $A\!\subseteq\!\Si_A^\om$ is an $\om$-power of a \la $L_A$, i.e., $A\! =\! L_A^\infty$, then $A^\approx$ is also an 
$\om$-power, i.e., there exists a (finitary) language $E_A$ such that $A^\approx\! =\! E_A^\infty$. Moreover, if the language $L_A$ is in the class $OCL(k)$ for some natural number $k$, then the language $E_A$ can be found in the class $OCL(k\! +\! 1)$. 
\end{cor}

\proo  Let $h\! :\!\Si_A\!\rightarrow\! 2^{\Si^{<\omega}}$ be the substitution defined by $a\!\mapsto\! L_3a$, where $L_3$ is the context-free  language defined above. Then it is easy to see that  now $A^\approx$ is obtained by substituting in $A$ the language
$L_3a$ for each letter $a\!\in\!\Si_A$. Thus $E_A\! =\! h(L_A)$ satisfies the statement of the theorem.\hfill{$\square$}\medskip

 The following result was proved in \cite{Fin01a}. 

\begin{thm}\label{pin} 
For each natural number $n\!\geq\! 1$, there is a context-free language $P_n$ in the subclass of iterated counter languages such that $P_n^\infty$ is ${\bf \Pi}_n^0$-complete.
\end{thm}

\proo
 Let $B_1\! =\!\{\sigma\!\in\!\{ {\bf 0},{\bf 1}\}^\om\mid\fa i\!\in\!\om ~~\sigma (i)\! =\! {\bf 0}\}\! =\! {\bf 0}^\infty$. $B_1$ is a ${\bf \Pi}_1^0$-complete set of the form $P_1^\infty$ where $P_1$ is the singleton containing only the word $({\bf 0})$. Note  that $P_1\! =\! {\bf 0}$ is a regular  language, hence in the class $OCL(0)$.\medskip

 Let then $B_2\! =\!\mathbb{P}_\infty$ be the well known ${\bf \Pi}_2^0$-complete regular $\om$-language. Note that 
$B_2\! =\! ({\bf 0}^{<\om}{\bf 1})^\infty$. Let $P_2\! :=\! {\bf 0}^{<\om}{\bf 1}$. Then $P_2$ is  a regular  language, hence in the class $OCL(0)$.\medskip
 
 We can now use iteratively  Corollary \ref{presop} to end the proof  \hfill{$\square$}\medskip

 Note that $P_1$ and $P_2$ are regular, hence accepted by some (real-time deterministic) finite automata (without any counter).
 On the other hand, the language $P_3$ is  accepted by a  one-counter automaton.  Notice that the $\om$-powers of regular languages are regular $\om$-languages, and thus are boolean combination of ${\bf \Pi}_2^0$-sets, hence ${\bf \Delta}_3^0$-sets. Therefore there are no ${\bf \Pi}_3^0$-complete or ${\bf \Si}_3^0$-complete (or even higher in the Borel hierarchy) $\om$-powers of regular languages. 
 \medskip 
 
 For the classes ${\bf \Sigma}_n^0$,  we first give an example of a ${\bf \Sigma}_n^0$-complete $\om$-power for 
$n\! =\! 1, 2$. Consider the finitary language 
$S_1\! :=\!\{ s\!\in\! 2^{<\omega}\mid 0\!\subseteq\! s\mbox{ or }\exists k\!\in\!\om ~~10^k1\!\subseteq\! s\}$ which is regular. Then the $\om$-power 
$S_1^\infty\! =\! 2^\omega\!\setminus\!\{ 10^\infty\}$ is open and not closed, and thus ${\bf \Sigma}_1^0$-complete.\medskip 
 
 Using another modification of the operation of exponentiation, we proved in \cite{Fink-Lec2} that there exists a one counter language 
$L\!\subseteq\! 2^{<\om}$ such that $L^\infty$ is ${\bf \Si}_2^0$-complete. It is enough to find a finitary language  
$S_2\!\subseteq\! 3^{<\om}$, where $3 = \{{\bf 0}, {\bf 1}, {\bf 2}\}$. We set, for $j\!\in\! 3$ and $s\!\in\! 3^{<\om}$, 
$$n_j(s)\! :=\!\mbox{Cardinality}(\{ i\! <\! |s|\mid s(i)\! =\! j\} )\mbox{,}$$ 
$T\! :=\!\{\alpha\!\in\! 3^{\leq\om}\mid\forall l\! <\! 1\! +\!\vert\alpha\vert ~~n_{\bf 2}(\alpha\vert l)\!\leq\! n_{\bf 1}(\alpha\vert l)\}$. We inductively define, for $s\!\in\! T\cap 3^{<\om}$, a ``back space" sequence $s^{\hookleftarrow}\!\in\! 2^{<\om}$ as follows:
$$s^{\hookleftarrow}\!  :=\!
\left\{\!\!\!\!\!\!\!
\begin{array}{ll}
& \emptyset\mbox{ if }s\! =\!\emptyset\mbox{,}\cr
& t^{\hookleftarrow}\varepsilon\mbox{ if }s\! =\! t\varepsilon\mbox{ and }\varepsilon\!\in\! 2\mbox{,}\cr
& t^{\hookleftarrow}\mbox{, except that its last {\bf 1} is replaced with {\bf 0}, if }s\! =\! t{\bf 2}.
\end{array}
\right.$$

 We then set $E\! :=\! {\bf 0}\cup\big\{ s\!\in\! T\cap 3^{<\om}\!\setminus\!\{\emptyset\}\mid n_{\bf 2}(s)\! =\! n_{\bf 1}(s)\mbox{ and }
{\bf 1}\!\subseteq\!\big( s\vert (|s|\! -\! 1)\big)^{\hookleftarrow}\big\}$, and 
$$E^*\! :=\!\{ {^\frown}_{i<l}~s_i\!\in\! 3^{<\om}\mid l\!\in\!\om\mbox{ and }\forall i<l~~s_i\!\in\! E\} .$$ 
We put $S_2\! :=\! E\cup\{ {^\frown}_{j\leq k}~(c_j{\bf 1})\!\in\! 3^{<\om}\mid k\!\in\!\om\mbox{ and }
( k\! =\! 0\Rightarrow c_0\!\not=\!\emptyset )\mbox{ and }\forall j\!\leq\! k~~c_j\!\in\! E^*\big\}$, and $S_2^\infty$ is $\boratwo$-complete. Note that $S_2$ is accepted by a one-counter automaton.\medskip
    
 Finally, we  recently proved in \cite{Fink-Lec3} the following result giving some complete $\omega$-powers of a one-counter language, for any Borel class of finite rank. 

\begin{thm} \label{main} Let $n\!\geq\! 1$ be a natural number.\smallskip

(a) There is a finitary language $P_n$ which is accepted by a one-counter automaton and such that the $\om$-power $P_n^\infty$ is ${\bf\Pi}_n^0$-complete.\smallskip

(b) There is a finitary language $S_n$ which is accepted by a one-counter automaton and such that the $\om$-power $S_n^\infty$ is ${\bf\Sigma}_n^0$-complete.\smallskip

Moreover, for any given integer $n\!\geq\! 1$, one can effectively construct some one-counter automata accepting such finitary languages $P_n$ and $S_n$ (here a construction is effective if there is an algorithm allowing it).\end{thm}

\subsection{$\!\!\!\!\!\!$ Borel $\om$-powers of infinite rank}\indent

  A first example of  an $\om$-power  which is a Borel set of infinite rank was obtained in \cite{Fin04-FI}. The idea was to  iterate  the operation $L\!\ra\! L^\approx$,   using  an infinite number of erasers. \medskip
  
 We can first iterate $k$ times this operation $A\!\ra\! A^\approx$. More precisely, we define, for a set $A\subseteq \Si^{\om}$, where $\Si$ is a finite alphabet. \medskip
 
\noindent - $A_k^{\approx .0}\! :=\! A$,\smallskip

\noindent - $A_k^{\approx .1}\! :=\! A^\approx$,\smallskip

\noindent - $A_k^{\approx .2}\! :=\! (A_k^{\approx .1})^\approx$,

\noindent\ldots \ldots \ldots \ldots \ldots \ldots \ldots \ldots\smallskip

\noindent - $A_k^{\approx .(k)}\! :=\! (A_k^{\approx .(k-1)})^\approx$,\medskip

\noindent where we apply $k$ times the operation $A\ra A^\approx$ with different new letters $\tla_{k}$, $\tla_{k-1}$, \ldots ,$\tla_3$, 
$\tla_2$, $\tla_1$, in such a way that we successively have\medskip
  
\noi $A_k^{\approx .0}\! =\! A\subseteq \Si^{\om}$,\smallskip

\noi $A_k^{\approx .1}\!\subseteq\! (\Si\cup\{\tla_k\})^{\om}$,\smallskip
   
\noi $A_k^{\approx .2}\!\subseteq\! (\Si\cup\{\tla_k, \tla_{k-1}\})^{\om}$,\smallskip

\noi\ldots \ldots \ldots \ldots \ldots \ldots \ldots \ldots \ldots \ldots \ldots \ldots \ldots \ldots\smallskip

\noi $A_k^{\approx .(k)}\!\subseteq\! (\Si \cup\{\tla_k, \tla_{k-1}, \ldots , \tla_1\})^{\om}$.\medskip

\noi and we set  ~~~~~~  $A^{\approx .(k)} = A_k^{\approx .(k)}$.\medskip

 Note that the choice of the erasers $\tla_{k}$, $\tla_{k-1}$, \ldots , $\tla_2$, $\tla_1$ in this precise order is important in the proof in  \cite{Fin04-FI}.\medskip
  
 We can now describe the operation $A\!\ra\! A^{\approx .(k)}$ in a manner similar to the case of the operation $A\!\ra\! A^{\approx}$, using the notion of a substitution.\medskip
 
 Let $T_k \subseteq (\Si\cup \{\tla_k, \tla_{k-1}, \ldots , \tla_1\})^{< \om}$ be the language containing the finite words $u$ over the alphabet $\Si\cup \{\tla_k, \tla_{k-1}, \ldots , \tla_1\}$ such that one gets the empty word after applying to $u$ the successive erasing operations with the erasers $\tla_1$, $\tla_2$, \ldots , $\tla_{k-1}$, $\tla_{k}$. More precisely, $u\!\in\! T_k$  if  when we start with $u$, we evaluate $\tla_1$ as an eraser, and obtain $u_1=u^{\tla_1}$  (following Definition \ref{approx}, i.e., every occurrence of a symbol $\tla_1$ does erase a letter of $\Si$ or an eraser $\tla_i$ for $i\! >\! 1$). Then we start again with $u_1$, this time we evaluate $\tla_2$ as an eraser, which yields $u_2=u_1^{\tla_2}$, and so on. When there is no more symbol $\tla_i$ to be evaluated, then there remains 
$u_k\!\in\!\Si^{<\om}$. By definition, $u\!\in\! T_k$ if and only if $u_k\! =\!\lambda$. It is  easy to see that $T_k$ is a context free language belonging to  the subclass of iterated counter languages.\medskip

 Now let $h_k$  be the substitution $\Si\!\ra\! 2^{((\Si \cup\{\tla_k, \tla_{k-1}, \ldots , \tla_1\})^{<\om})}$ defined by 
$h_k(a)\! :=\! {L_k}^\frown a$ for every letter $a\in \Si$. It holds that $A^{\approx .(k)}\! =\! h_k(A)$, for every $A \subseteq \Sio$.\medskip

 We now set $\Sigma=\{{\bf 0}, {\bf 1}\}$. Consider now the $\om$-language $B_2\! :=\! ({\bf 0}^{<\om}{\bf 1})^\infty\! =\! P_2^\infty$, where $P_2$ is the language ${\bf 0}^{<\om}{\bf 1}$. $B_2$ is ${\bf \Pi}_2^0$-complete. Then, as in the proof of Theorem \ref{pin}, 
$h_{p}(P_2^\infty)=(h_{p}(P_2))^\infty$ is a ${\bf \Pi}_{p+2}^0$-complete set, for each integer $p\!\geq\! 1$.\medskip

 On the other hand,  the languages $T_k$, for $k\geq 1$, form a sequence which is strictly increasing for the inclusion relation: 
$$T_1\!\subsetneq\! T_2\!\subsetneq\! T_3\!\subsetneq\!\ldots\!\subsetneq\! T_i\!\subsetneq\! T_{i+1}\ldots$$
In order to construct an $\om$-power which is Borel of infinite rank, the first idea is to substitute the language 
$\bigcup_{k\geq 1}{L_k} ^\frown a$ to each letter $a\!\in\!\Si\! =\!\{{\bf 0},{\bf 1}\}$ in the language $P_2^\infty$. But this way we would get a language over the {\it infinite} alphabet  $\Si \cup\{\tla_1, \tla_2, \tla_3, \ldots\}$. In order to obtain a finitary language over a {\it finite} alphabet, every eraser $\tla_j$ can be coded by a finite word $\alpha.\beta^j.\alpha$  over the alphabet $\{\alpha, \beta\}$, where 
$\alpha$ and $\beta$ are two new letters. \medskip 

 One defines the substitution $\vp_k\!:\! (\Si\cup\{\tla_1,\ldots ,\tla_k\})^{<\om}\!\ra\! 2^{(\Si\cup\{\alpha, \beta\})^{<\om}}$ by 
$\vp_p(c)\! :=\!\{ c\}$ for each $c\!\in\!\Si$ and $\vp_k(\tla_j)\! =\!\{\alpha.\beta^j.\alpha\}$ for each integer $j\!\in\! [1,k]$. Now let  
$\mathcal{L}\! :=\!\bigcup_{k\geq 1}~\vp_k(T_k)$, and $h\! :\!\Si\!\ra\! 2^{((\Si \cup \{\alpha, \beta\})^{<\om})}$ be the substitution defined by $h(a)\! :=\!\mathcal{L}^\frown a$, for each $a\in \Si$.

\begin{thm}
\label{bor-inf-rank} Let $P_2\! :=\! {\bf 0}^{<\om}{\bf 1}$. Then the $\om$-power 
$( h(P_2) )^\infty\!\subseteq\!\{ {\bf 0},{\bf 1},\alpha,\beta\}^\om$ is a Borel set of infinite rank.\end{thm}

The language $( h(P_2) )$ is a simple recursive language but it is not context-free. Later, with a modification of the construction, and using a coding of an infinity of erasers previously defined in \cite{Fin03b}, Finkel and Duparc got a context-free language $W$ such that 
 $W^\infty$ is a Borel set of infinite rank \cite{Fin-Dup06}.

\begin{thm}
There exists a context-free finitary language $W \subseteq \Gamma^{< \om}$, where $\Gamma$ is a finite alphabet, such that $W^\infty$ is a Borel set of 
infinite rank. Moreover $W^\infty$  is above the class ${\bf \Delta}_\om^0$. 
\end{thm}

The coding of the infinity of erasers $\tla_n$ is given by $\Phi(\tla_n)=\alpha B^n C^n D^n E^n \beta$ with new letters $\alpha, B, C, D, E, \beta$. Actually the pushdown automaton constructed in order to accept the language $W$ must  be able to read 
the number $n$ identifying the eraser four times. \medskip

 The $\om$-power $W^\infty$ is above the class ${\bf \Delta}_\om^0$, i.e., it is not in the Borel class ${\bf \Delta}_\om^0$. Note that the 
$\om$-power $( h(P_2) )^\infty$ was actually also above the class ${\bf \Delta}_\om^0$ but this was not shown in \cite{Fin04-FI}.  We give the argument in this latter case, where the language  $h(P_2)$ is simpler than $W$. This follows from the fact that 
$(( h(P_2) )^\infty)^\approx $ is Wadge equivalent to $( h(P_2) )^\infty$, which is due to the precise way we ordered the erasers, as  described above. On the other side the operation $A\!\rightarrow\! A^{\approx}$ is strictly increasing for the Wadge ordering inside 
${\bf\Delta}^0_{\omega}$ (see \cite{Duparc01}). This implies that $( h(P_2) )^\infty$, and also $W^\infty$, are not in the class 
${\bf\Delta}^0_{\omega}$.\medskip 

 Note that the language $W$ is context-free but it cannot be accepted by a one-counter automaton.

\subsection{$\!\!\!\!\!\!$ Non-Borel $\om$-powers which are even ${\bf \Sigma}_1^1$-complete}\indent

 A first  example of  language $L$ such that $L^\infty$ is  not Borel, and even ${\bf \Si}_1^1$-complete, was obtained in  \cite{Fin03a}. 
It turned out that the language $L$ may be described in a very simple way. Surprisingly  it is actually  accepted by a one-counter automaton. It was obtained   via a coding of infinite labelled  binary trees. We now recall the construction of this language $L$ using the notion of a substitution.\medskip 

 Let $d$ be a letter not in $2$ and 
$D\! :=\!\{ ~u\!\cdot\! d\!\cdot\! v\mid u,v\!\in\! 2^{<\om}\mbox{ and }|v|\! =\! 2|u|\mbox{ or }|v|\! =\! 2|u|\! +\! 1~\}$. It is easy to see that the language $D\!\subseteq\! (2\cup\{ d\})^{<\om}$ is a context-free language accepted by a one-counter automaton.\medskip 

 Let $g\! :\!\Si\!\ra\! 2^{(2 \cup \{d\})^{<\om}}$ be the substitution defined by $g(a)\! =\! a\!\cdot\! D$. Since 
$W\! :=\!\{{\bf 0}\}^{<\om}\!\cdot\! {\bf 1}$ is a regular language, $L\! :=\! g(W)$ is a context-free language and it is accepted by a one-counter automaton. Moreover, it is proved in \cite{Fin03a} that $(g(W))^\infty$ is ${\bf \Si}^1_1$-complete, and thus non-Borel. This is done by reducing to this $\om$-language a well-known example of a ${\bf \Si}^1_1$-complete set: the set of infinite  binary trees labelled in the alphabet $2$ which have an infinite branch in the ${\bf \Pi}^0_2$-complete set $W^\infty$. 

\section{$\!\!\!\!\!\!$ Classical and effective complexity of the $\omega$-powers}\indent

 In \cite{Fin-Lec}, we prove that there are some $\om$-powers of any Borel rank. More precisely, Theorem 2 in \cite{Fin-Lec} is as follows.

\begin{thm} \label{rk} Let $\xi\!\geq\! 1$ be a countable ordinal.\smallskip

(a) There is a finitary language $P_\xi\!\subseteq\! 2^{<\om}$ such that the $\om$-power $P_\xi^\infty$ is ${\bf\Pi}_\xi^0$-complete.\smallskip

(b) There is a finitary language $S_\xi\!\subseteq\! 2^{<\om}$ such that the $\om$-power $S_\xi^\infty$ is ${\bf\Sigma}_\xi^0$-complete.\end{thm}

 In fact, we provide a general method proving this when $\xi\!\geq\! 3$. Examples of such finitary languages were 
 given in Section \ref{Si0N-Pi0N} when $\xi\!\leq\! 2$.\medskip

 We now turn to the general case. Let ${\bf\Gamma}$ be a class of sets of the form $\boraxi$ or $\bormxi$, with $\xi\!\geq\! 3$. Fix a $\bf\Gamma$-complete set $B\!\subseteq\! 2^\omega$, so that $B\!\in\! {\bf\Pi}^0_{\xi +1}$. A result due to Kuratowski provides a closed subset $C$ of $\om^\om$ and a continuous bijection $f\! :\! C\!\rightarrow\! B$ with the property that $f^{-1}$ is $\boraxi$-measurable (i.e., $f[O]$ is a $\boraxi$ subset of $B$ if $O$ is an open subset of $C$, see \cite{Kur66}). This result is a level by level version of a result, due to Lusin and Souslin, asserting that every Borel subset of $2^\om$ is the image of a closed subset of $\om^\om$ by a continuous bijection. By Proposition 11 in \cite{Lecomte05}, it is enough to find a finitary language $A\!\subseteq\! 4^{<\om}$, where $4\! :=\!\{ {\bf 0},{\bf 1},{\bf 2},{\bf 3}\}$, such that $A^\infty$ is  ${\bf\Gamma}$-complete.
 
\vfill\eject
 
  The  language $A$ will be made of two pieces: $A\! =\!\mu\cup\pi$. The set $\pi$ will code $f$, and $\pi^\infty$ will look like $B$ on some compact sets $K_{N,j}$. Outside this countable family of compact sets we will hide $f$, so that $A^\infty$ will be the simple set $\mu^\infty$.\medskip

 The Lusin-Souslin theorem has been used by Arnold in \cite{Arnold83} to prove that every Borel subset of $\Si^\om$, where $\Si$ is a finite alphabet, is accepted by a non-ambiguous finitely branching transition system with a B\"uchi acceptance condition, and our first idea was to code the behaviour of such a transition system.

\begin{defi} A {\bf B\"uchi\ transition\ system} is a 5-tuple $\mathcal{T}\! =\! (Q,\Si ,q_0,\Delta ,F)$, where $Q$ is a (possibly infinite) countable set of states, $\Sigma$ is a finite input alphabet, $q_0\!\in\! Q$ is the initial state, $\Delta\!\subseteq\! Q\!\times\!\Si\!\times\! Q$ is the transition relation, and $F\!\subseteq\! Q$ is the set of final states.\smallskip
  
 Let $\sigma\! =\! a_0a_1\ldots$ be an $\om$-word over $\Si$. An $\om$-sequence of states $r\! =\! (t_i)_{i\in\omega}$ is called a {\bf run of $\mathcal{T}$ on $\sigma$} if
\begin{enumerate}
\ite[(1)] $t_0\! =\! q_0$,

\ite[(2)] for each $i\!\in\!\omega$, $\big( t_i,\sigma (i),t_{i+1}\big)\!\in\!\Delta$.
\end{enumerate}

 The run $r$ is said to be {\bf accepting} when $t_i\!\in\! F$ for infinitely many $i$'s. The transition system $\mathcal{T}$ is said to be\smallskip
 
- {non}-{\bf ambiguous} if each infinite word $\sigma\!\in\!\Si^\om$ has at most one accepting run by $\mathcal{T}$,\smallskip

- {\bf finitely\ branching} if for each state $q\!\in\! Q$ and each $a\!\in\!\Si$, there are only finitely many states $q'$ such that 
$(q, a, q')\!\in\!\Delta$.\smallskip

 The \ol~accepted by $\mathcal{T}$ is 
\begin{center}
$A(\mathcal{T})\! :=\!\{\sigma\!\in\!\Si^\om\mid\mbox{ there exists an accepting  run }r~{ of }~\mathcal{T}\mbox{ on }\sigma\}$.
\end{center}
\end{defi}

 We will code the behaviour of a transition system coming from $f$.\medskip
 
\noindent - The set of states is $Q\! :=\!\{ (s,t)\!\in\! 2^{<\om}\!\times\! 2^{<\om}\mid |s|\! =\! |t|\}$, which is countably infinite. We enumerate $Q$ as follows. We start with $q_0\! :=\! (\emptyset ,\emptyset )$. Then we put the sequences of length 1 of elements of $2\!\times\! 2$, in the lexicographical ordering: $q_1\! :=\! (0,0)$,  $q_2\! :=\! (0,1)$, $q_3\! :=\! (1,0)$, $q_4\! :=\! (1,1)$. Then we put the 16 sequences of length 2: $q_5\! :=\! (0^2,0^2)$, $q_6\! :=\! (0^2,01)$, ... And so on.\medskip

 We will sometimes use the coordinates of $q_n\! :=\! (q_n^0 ,q_n^1)$. We put $M_j\! :=\!\Si_{i<j}~4^{i+1}$. Note that the sequence 
$(M_j)_{j\in\om}$ is strictly increasing, and that $q_{M_j}$ is the last sequence of length $j$ of elements of $2\!\times\! 2$. We define, for $N,j\!\in\!\om$ with $N\!\leq\! M_j$, the compact set 
$$K_{N,j}\! :=\!\{ ~{\bf 2}^N({^\frown}_{i\in\om}~m_i~{\bf 2}^{M_{j+i+1}}~{\bf 3}~{\bf 2}^{M_{j+i+1}})\!\in\! 4^\om\mid
\forall i\!\in\!\om ~~m_i\!\in\! 2~\} .$$
- The input alphabet is $2$.\medskip

\noindent - The initial state is $q_0\! :=\! (\emptyset ,\emptyset )$.\medskip

\noindent - If $m\!\in\! 2$ and $n,p\!\in\!\om$, then we write $n\!\xrightarrow{m}\! p$ if $q_n^0\!\subseteq\! q_p^0$ and $q_p^1\! =\! q_n^1m$. As $f$ is continuous on $C$, the graph $\mbox{Graph}(f)$ of $f$ is a closed subset of $C\!\times\! 2^\om$. As $C$ is a closed subset of 
$\mathbb{P}_\infty$, $\mbox{Graph}(f)$ is also a closed subset of $\mathbb{P}_\infty\!\times\! 2^\om$. So there is a closed subset $P$ of $2^\om\!\times\! 2^\om$ with the property that 
$$\mbox{Graph}(f)\! =\! P\cap (\mathbb{P}_\infty\!\times\! 2^\om ).$$ 
We identify $2^\om\!\times\! 2^\om$ with $(2\!\times\! 2)^\om$, i.e., we view $(\beta ,\alpha )$ as 
$\big(\beta (0),\alpha (0)\big) ,\big(\beta (1),\alpha (1)\big) , $...

\vfill\eject

 By Proposition 2.4 in \cite{Kechris94}, there is $R\!\subseteq\!(2\!\times\! 2)^{<\om}$, closed under initial segments, such that 
$$P\! =\!\{ (\beta ,\alpha )\!\in\! 2^\om\!\times\! 2^\om\mid\forall k\!\in\!\om ~~(\beta ,\alpha )\vert k\!\in\! R\}\mbox{;}$$ 
note that $R$ is a tree whose infinite branches form the set $P$. In particular, we get 
$$(\beta ,\alpha )\!\in\!\mbox{Graph}(f)\Leftrightarrow \beta\!\in\!\mathbb{P}_\infty\mbox{ and }
\forall k\!\in\!\om ~~(\beta ,\alpha )\vert k\!\in\! R.$$
The transition relation $\Delta\!\subseteq\! Q\!\times\! 2\!\times\! Q$ is given by 
$(q_n,m,q_p)\!\in\!\Delta\Leftrightarrow n\!\xrightarrow{m}\! p$, for $m\!\in\! 2$ and $n,p\!\in\!\om$.\medskip

\noindent - The set of final states is $F\! :=\!\{ (t,s)\!\in\! R\mid t\!\not=\!\emptyset\mbox{ and }t(|t|\! -\! 1)\! =\! 1\}$. Note that $F$ is simply the set of pairs $(t,s)\!\in\! R$ such that the last letter of $t$ is a 1.\medskip

 Recall that a run of $\mathcal{T}$ is said to be B\"uchi accepting if  final states occur infinitely often during this run. Then the set of 
$\om$-words over the alphabet 2 which are accepted by the transition system $\mathcal{T}$ from the initial state $q_0$ with B\"uchi acceptance condition is exactly the Borel set $B$.\medskip

 We are now ready to define the finitary language $\pi$. We set
$$\pi\! :=\!\left\{
\begin{array}{ll}
s\!\in\! 4^{<\om}\mid\exists j,l\!\in\!\om\!\!\! 
& \exists (m_i)_{i\leq l}\!\in\! 2^{l+1}~~\exists (n_i)_{i\leq l},(p_i)_{i\leq l},(r_i)_{i\leq l}\!\in\!\om^{l+1}\cr\cr
& n_0\!\leq\! M_j\cr
& ~~~~~\mbox{ and }\cr
& \forall i\!\leq\! l~~n_i\!\xrightarrow{m_i}\! pi\mbox{ and }p_i\! +\! r_i\! =\! M_{j+i+1}\cr
& ~~~~~\mbox{ and }\cr
& \forall i\! <\! l~~p_i\! =\! n_{i+1}\cr
& ~~~~~\mbox{ and }\cr
& q_{p_l}\!\in\! F\cr
& ~~~~~\mbox{ and }\cr
& s\! =\! {^\frown}_{i\leq l}~{\bf 2}^{n_i}~m_i~{\bf 2}^{p_i}~{\bf 2}^{r_i}~{\bf 3}~{\bf 2}^{r_i}
\end{array}
\right\} .$$

 We are also ready to define $\mu$. The idea is that an infinite sequence containing a word in $\mu$ cannot be in the union of the 
$K_{N,j}$'s. We set
$$\begin{array}{ll} 
& \mu_0\!:=\!\left\{      
\begin{array}{ll}
s\!\in\! 4^{<\om}\mid\exists l\!\in\!\om ~~\exists (m_i)_{i\leq l+1}\!\in\! 2^{l+2}\!\!\!
& \exists N\!\in\!\om  ~~\exists (P_i)_{i\leq l+1},(R_i)_{i\leq l+1}\!\in\!\om^{l+2}\cr\cr
& \forall i\!\leq\! l\! +\! 1~~\exists j\!\in\!\om ~~P_i\! =\! M_j\cr
& ~~~~~\mbox{ and }\cr
& P_l\!\not=\! R_l\cr
& ~~~~~\mbox{ and }\cr
& s\! =\! {\bf 2}^N({^\frown}_{i\leq l+1}~m_i~{\bf 2}^{P_i}~{\bf 3}~{\bf 2}^{R_i}) 
\end{array}
\right\}\mbox{,}\cr\cr
& \mu_1\!:=\!\left\{      
\begin{array}{ll}
s\!\in\! 4^{<\om}\mid\exists l\!\in\!\om ~~\exists (m_i)_{i\leq l+1}\!\in\! 2^{l+2}\!\!\!
& \exists N\!\in\!\om  ~~\exists (P_i)_{i\leq l+1},(R_i)_{i\leq l+1}\!\in\!\om^{l+2}\cr\cr
& \forall i\!\leq\! l\! +\! 1~~\exists j\!\in\!\om ~~P_i\! =\! M_j\cr
& ~~~~~\mbox{ and }\cr
& \exists j\!\in\!\om ~~(P_l\! =\! M_j\mbox{ and }P_{l+1}\! \neq \! M_{j+1})\cr
& ~~~~~\mbox{ and }\cr
& s\! =\! {\bf 2}^N({^\frown}_{i\leq l+1}~m_i~{\bf 2}^{P_i}~{\bf 3}~{\bf 2}^{R_i}) 
\end{array}
\right\}\mbox{,}
\end{array}$$
and $\mu\! :=\!\mu_0\cup\mu_1$. Recall that $A\! =\!\mu\cup\pi$.

\vfill\eject

 We just described how to get the finitary languages in the statement of Theorem \ref{rk}. For the other Borel classes $\borxi$, only 
$\borone$ is a Wadge class, and $A\! :=\!\{ s\!\in\! 2^{<\om}\mid 0\!\subseteq\! s\mbox{ or }1^2\!\subseteq\! s\}$ has the property that 
$A^\infty\! =\! 2^\om\!\setminus\! N_{10}$ is $\borone$-complete (see \cite{Fink-Lec2}). In \cite{Fink-Lec2}, we provide some complete sets for some other Wadge classes of Borel sets, in fact some dual classes of classes of differences of $\boraxi$ sets (see also \cite{Lecomte05}). It is worth noting that Theorem \ref{rk} may seem to indicate that $\om$-powers can be arbitrarily complex, but its proof uses closure properties of the classes of the Borel hierarchy that are not shared by all the Wadge classes of Borel sets, such as the closure by finite unions. The extension of Theorem \ref{rk} to all Wadge classes of Borel sets is an open problem.\medskip 

 An important result in \cite{Fink-Lec2} shows that Theorem \ref{rk} is as effective as it can be, in the context of effective descriptive set theory. In order to state it, we must recall some notions about this theory. Effective descriptive set theory is based on the notion of a recursive function.  A function from $\om^k$ to $\om$ is said to be {\bf recursive} if it is total and computable.  By extension, a relation is called {\bf recursive} if its characteristic function is recursive.

\begin{defi}  
  A {\bf recursive presentation} of a topological space $X$ is a pair
  $\big( (x_n)_{n\in\om},d\big)$ such that
  \begin{enumerate}
  \item $(x_n)_{n\in\om}$ is dense in $X$,
    
  \item $d$ is a compatible complete distance on $X$ such that the following
  relations $P$ and~$Q$ are recursive:
  \begin{align*}
    P(i,j,m,k) & \iff d(x_i,x_j) \leq \frac{m}{k+1}, \\
    Q(i,j,m,k) & \iff d(x_i,x_j) < \frac{m}{k+1}.
  \end{align*}
\end{enumerate}
A topological space $X$ is {\bf recursively presented} if it is given with a recursive presentation of it.
\end{defi}

  Note that every recursively presented space is {\bf Polish} (i.e., separable and completely metrizable). For example, one can check that the spaces $\om$ and $\Si^\om$ have a recursive presentation. Moreover, a product of two recursively presented spaces has a recursive presentation.\medskip
  
 Note that the formula $(p,q)\!\mapsto\! 2^p(2q\! +\! 1)\! -\! 1$ defines a recursive bijection $\om^2\!\rightarrow\!\om$. One can check that the coordinates of the inverse map are also recursive. They will be denoted $n\!\mapsto\! (n)_0$ and $n\!\mapsto\! (n)_1$ in the sequel.  These maps will help us to define some of the basic effective classes.

\begin{defi} \label{eff} Let $\big( (x_n)_{n\in\om},d\big)$ be a recursive presentation of a topological space $X$.

  \begin{enumerate}
  
  \item We fix a countable basis of $X$: $B(X,n)$ is the open ball $B_d(x_{(n)_0},\frac{( (n)_1)_0}{( (n)_1)_1+1})$.

  \item A subset $S$ of $X$ is {\bf semirecursive}, or {\bf effectively
      open} (denoted $S\!\in\!\Boraone$) if
    $$S\! =\!\bigcup_{n\in\om}{B\big( X,f(n)\big)}\mbox{,}$$ for some recursive
    function $f$.

  \item If $n\!\geq\! 1$ is a natural number, then 
$\Bormn$ is the class of complements of $\Boran$ sets. We say that 
$B\!\in\!\Boranpo$ if there is $C\!\in\!\Bormn (\omega\!\times\! X)$ such that 
$B\! =\!\exists^\omega C\! :=\!\{ x\!\in\! X\mid\exists i\!\in\!\omega\ (i,x)\!\in\! C\}$. We also set 
$\Born\! :=\!\Boran\cap\Bormn$.

\vfill\eject

  \item A subset $S$ of
    $X$ is {\bf effectively analytic} (denoted $S\!\in\!\Ana$) if there
    is a $\Bormone$ subset $C$ of $X\!\times\!\om^\om$ such
    that $S\! =\!\mbox{proj}_X[C]\! :=\!\{ x\!\in\! X\mid\exists\alpha\!\in\!\om^\om ~~(x,\alpha)\!\in\! C\}$. A subset $S$ of $X$ is 
    {\bf effectively co-analytic} (denoted $S\!\in\!\Ca$) if its complement $\neg S$ is effectively analytic, and {\bf effectively Borel} if it is in 
    $\Ana$ and $\Ca$ (denoted $S\!\in\!\Borel$). We also set ${\it\Sigma}^1_2\! :=\!\{\exists^{\om^\om}C\mid C\!\in\!\Ca\}$, 
    ${\it\Pi}^1_2\! :=\!\check {\it\Sigma}^1_2$ and ${\it\Delta}^1_2\! :=\! {\it\Sigma}^1_2\cap {\it\Pi}^1_2$. 
    
\item We will consider the {\bf relativized classes}: if $Y$ is a recursively presented space and $y \in  Y$, then we say
  that $A\!\subseteq\! X$ is in $\Ana (y)$ if there is
  $S\!\in\!\Ana (Y\!\times\! X)$ such that 
  $$A\! =\! S_y\! :=\!\{ x\!\in\! X\mid (y,x)\!\in\! S\} .$$ 
  The class $\Ca (y)$ is defined similarly. We also set  $\Borel (y) := \Ana (y)\cap\Ca (y)$.
  
\item Let $\gamma\!\in\!\om^\om$. We say that $\gamma\!\in\!\Boraone$ if 
$\{ k\!\in\!\omega\mid\gamma\!\in\! B(\omega^\omega ,k)\}\!\in\!\Boraone (\omega)$. A countable ordinal $\xi$ is a {\bf recursive ordinal} if there is $\gamma\!\in\!\Boraone$ coding a well-ordering on $\om$ of order type $\xi$ .

\item There is a {\bf good parametrization} in $\Boraone$ for $\boraone$ (see 3E.2, 3F.6 and 3H.1 in \cite{Moschovakis80}). This means that there is a system of sets $G^{\Boraone ,Y}\!\in\!\Boraone (\omega^\omega\!\times\! Y)$ such that, for each recursively presented space $Y$ and for each $P\!\subseteq\! Y$, 
$$\begin{array}{ll}
P\!\in\!\boraone & \!\!\Leftrightarrow\  \ 
\exists\gamma\!\in\!\omega^\omega\ \ P\! =\! G^{\Boraone ,Y}_{\gamma}
\hbox{\it ,}\cr
P\!\in\!\Boraone & \!\!\Leftrightarrow \ \ 
\exists\gamma\!\in\!\Boraone\ \ P\! =\! G^{\Boraone ,Y}_{\gamma}.
\end{array}$$
Moreover, if $Z$ is a recursively presented space of type at most 1 
(i.e., a finite product of spaces equal to $\omega$, $\omega^\omega$ or 
$2^\omega$), and $Y$ is a recursively presented space, then there is 
$S^{Z,Y}_{\Boraone}\! :\!\omega^\omega\!\times\! Z\!\rightarrow\!\omega^\omega$ recursive such that 
$(\gamma ,z,y)\!\in\! G^{\Boraone ,Z\times Y}\ \ \Leftrightarrow\ \ \  
\big( S^{Z,Y}_{\Boraone}(\gamma ,z),y\big)\!\in\! G^{\Boraone ,Y}$ (here, by $S^{Z,Y}_{\Boraone}$ recursive we mean that the relation defined by $R(\gamma ,z,k)\Leftrightarrow S^{Z,Y}_{\Boraone}(\gamma ,z)\!\in\! B(\omega^\omega ,k)$ defines a $\Boraone$ subset of 
$\omega^\omega\!\times\! Z\!\times\!\omega$).

\item We can code the partial recursive functions. Let $Y$ be a recursively presented space, $f\! :\! X\!\rightarrow\! Y$ be a partial function, $D\!\subseteq\!\mbox{Domain}(f)$ and $P\!\subseteq\! X\!\times\!\omega$. Then $P$ {\bf computes} $f$ on $D$ if 
$$x\!\in\! D\ \ \Rightarrow\ \ \forall k\!\in\!\omega\ \ \big( f(x)\!\in\! B(Y,k)\ \Leftrightarrow\ (x,k)\!\in\! P\big) .$$
If $P$ is in $\Boraone$ and computes $f$ on $D$, then we say that $f$ is {\bf recursive on }$D$. This means that 
$f^{-1}\big( B(Y,k)\big)\!\in\!\Boraone$, uniformly in $k$.\smallskip

 We now define a partial function $U\! :\!\omega^\omega\!\times\! X\!\rightarrow\! Y$ by 
$$\begin{array}{ll}
U(\gamma ,x)\!\downarrow 
& \!\Leftrightarrow\ U(\gamma ,x)\mbox{\rm is\ defined}
\Leftrightarrow\exists y\!\in\! Y\ \forall 
k\!\in\!\omega\ \ \big( y\!\in\! B(Y,k)\Leftrightarrow 
(\gamma ,x,k)\!\in\! G^{\Boraone ,X\times\omega}\big)\mbox{\rm ,}\cr & \cr 
U(\gamma ,x) 
& \!\!\! :=\ \ \mbox{\rm the\ unique}\ y\!\in\! Y\ \hbox{\rm such that}\ \forall k\!\in\!\omega\ \ 
\big( y\!\in\! B(Y,k)\Leftrightarrow (\gamma ,x,k)\!\in\! G^{\Boraone ,X\times\omega}\big) .
\end{array}$$
Now let $\gamma\!\in\!\omega^\omega$. The function $\{\gamma\}^{X,Y}\! :\! X\!\rightarrow\! Y$ is defined by 
$\{\gamma\}^{X,Y}(x)\! :=\! U(\gamma ,x)$. Then a partial function $f\! :\! X\!\rightarrow\! Y$ is recursive on its domain if and only if there is $\gamma\!\in\!\Boraone$ such that $f(x)\! =\!\{\gamma\}^{X,Y}(x)$ when $f(x)$ is defined. More generally, the functions of the form 
$\{\gamma\}^{X,Y}$ are the partial continuous functions from a subset of $X$ into $Y$. In order to simplify the notation, we will write 
$\{\gamma\}$ instead of $\{\gamma\}^{X,Y}$ when $Y\! =\!\omega^\omega$.

\item We now define, by induction on the countable ordinal $\xi\!\geq\! 1$, the set $BC_{\xi}$ of Borel codes for $\boraxi$ as follows. If 
$\gamma\!\in\!\omega^\omega$, then we define $\gamma^*\!\in\!\omega^\omega$ by $\gamma^*(i)\! :=\!\gamma (i\! +\! 1)$. We set 
$$\begin{array}{ll}
BC_{1} & \!\!\! :=\{\ \gamma\!\in\!\omega^\omega\mid\gamma (0)\! =\! 0\ \}
\hbox{\rm ,}\cr 
BC_{\xi} & \!\!\! :=\Big\{\ \gamma\!\in\!\omega^\omega\mid\gamma (0)\! =\! 1
\mbox{ and }\forall i\!\in\!\omega\ \ \{\gamma^{*}\}(i)\!\downarrow
\mbox{ and }\{\gamma^{*}\}(i)\!\in\!\bigcup_{1\leq\eta <\xi}\ BC_{\eta}\ \Big
\}\ \hbox{\rm if}\ \xi\! \geq\! 2.\end{array}$$

 The set of Borel codes is $BC\! :=\!\bigcup_{1\leq\xi <\omega_{1}}\ BC_{\xi}$. We also set 
$BC^{*}\! :=\!\bigcup_{2\leq\xi <\omega_{1}}\uparrow\ BC_{\xi}$. We define $\rho^{X}\! :\! BC\!\rightarrow\!\borel (X)$ by induction:
$$\rho^{X}(\gamma ):=\left\{\!\!\!\!\!\begin{array}{ll}
& \!\!\!\bigcup_{i\in\omega}\ B\big( X,\gamma^{*}(i)\big)\ \ \hbox{\rm if}\ \ 
\gamma\!\in\! BC_{1}\hbox{\rm ,}\cr & \cr
& \!\!\!\bigcup_{i\in\omega}\ X\!\setminus\rho^{X}\big(\{\gamma^{*}\}(i)\big)\ \ 
\hbox{\rm if}\ \ \gamma\!\in\! BC^{*}.
\end{array}\right.$$ 
Clearly, $\rho^{X}[BC_{\xi}]\! =\!\boraxi (X)$, by induction on $\xi$.

\item We can now define the {\bf hyperarithmetical\ hierarchy}. Let $\xi\!\geq\! 1$ be a countable ordinal. Then 
$$\begin{array}{ll}
\Boraxi (X) & \!\!\!\! =\{\rho^{X}(\gamma )\mid\gamma\!\in\!\Boraone\cap BC_{\xi}\}
\hbox{\rm ,}\cr &\cr 
\Bormxi (X) & \!\!\!\! =\check\Boraxi (X)\hbox{\rm ,}\cr & \cr
\Borxi (X) & \!\!\!\! =\Boraxi (X)\cap\Bormxi (X).
\end{array}$$
This definition is compatible with the item 3.
   
\end{enumerate}
\end{defi}

 The crucial link between the effective classes and the classical corresponding classes is as follows: the class of analytic (resp.,
co-analytic, Borel) sets is equal to $\bigcup_{\alpha\in\om^\om}~\Ana (\alpha )$ (resp., $\bigcup_{\alpha\in\om^\om}~\Ca (\alpha )$,
$\bigcup_{\alpha\in\om^\om}~\Borel (\alpha )$). This allows to use effective descriptive set theory to prove results of classical type.
  
\begin{thm} \label{rkeff} Let $\xi\!\geq\! 1$ be a recursive ordinal.\smallskip

(a) There is a finitary language $P_\xi\!\subseteq\! 2^{<\om}$, that can be coded by a $\Borone$ subset of $\om$, such that the $\om$-power $P_\xi^\infty$ is in the effective class ${\it\Pi}_\xi^0$ but not in ${\bf\Sigma}_\xi^0$.\smallskip

(b) There is a finitary language $S_\xi\!\subseteq\! 2^{<\om}$, that can be coded by a $\Borone$ subset of $\om$, such that the $\om$-power $S_\xi^\infty$ is in the effective class ${\it\Sigma}_\xi^0$ but not in ${\bf\Pi}_\xi^0$.\end{thm}

\section{$\!\!\!\!\!\!$ Complexity of some sets of finitary languages related to the $\omega$-powers}\indent

 In \cite{Lecomte05}, the following question is raised. What is the topological complexity of the set of finitary languages whose associated $\om$-power is of a given level of complexity?\medskip
 
  This question arises naturally when we look at the characterizations of closed, $\bormtwo$ and open $\om$-powers obtained in \cite{Staiger97-2} (see Corollary 14 and Lemmas 25, 26). This leads to set, for a class of sets $\bf\Gamma$, 
$\mathcal{L}_{\bf\Gamma}\! :=\!\{ L\!\subseteq\! 2^{<\omega}\mid L^\infty\!\in\! {\bf\Gamma}\}$. It is proved in \cite{Lecomte05} (see Theorem 4) that $\mathcal{L}_{\{\emptyset\}}$ is $\bormone$-complete, $\mathcal{L}_{\check\{\emptyset\}}$ is $\boraone$-complete, and 

\begin{thm} \label{clopen} The set $\mathcal{L}_{\borone}$ is $\boratwo$-complete.\end{thm}

\vfill\eject

 For the next classes of the Borel hierarchy, it is proved in \cite{Lecomte05} that $\mathcal{L}_{\boraxi}$ are $\mathcal{L}_{\bormxi}$ are ${\bf\Sigma}^1_2$ (see Proposition 16). A consequence of Theorem \ref{rkeff} is that these sets are $\ca$-hard if $\xi\!\geq\! 3$ (see Corollary 6.4 in \cite{Fink-Lec2}). It is proved in \cite{Fin10} that for every integer $k\geq 2$ (respectively, $k\geq 3$) the set $\mathcal{L}_{{\bf \Pi}_{k+1}^0}$ (respectively,  $\mathcal{L}_{{\bf \Si}_{k+1}^0}$) is ``more complex'' than  the set $\mathcal{L}_{{\bf \Pi}_{k}^0}$ (respectively,  $\mathcal{L}_{{\bf \Si}_{k}^0}$), with respect to the Wadge reducibility. The following result is proved in \cite{Lecomte05,Fin10}.
 
 \begin{thm}
 The set $\mathcal{L}_{{\bf \Delta}_1^1}$ is in ${\bf \Sigma}_2^1 \setminus {\bf \Pi}^0_2$.
 \end{thm}

 Along similar lines, some other results of effective nature are available in \cite{Lecomte05,Fink-Lec2}. For instance, we set 
$\mathcal{L}_\Delta\! :=\!\{ L\!\subseteq\! 2^{<\omega}\mid L^\infty\!\in\!\Borel (L)\}$. The following is proved in \cite{Lecomte05} and \cite{Fink-Lec2}.
 
 \begin{thm} The following sets are co-analytic and not Borel.\smallskip
 
 (a) $\mathcal{L}_\Delta$,\smallskip
 
 (b) $\mathcal{L}_{\boraxi}\cap\mathcal{L}_\Delta$ ($\ca$-complete if $\xi\!\geq\! 3$),\smallskip
 
 (c) $\mathcal{L}_{\bormxi}\cap\mathcal{L}_\Delta$ if $\xi\!\geq\! 2$ ($\ca$-complete if $\xi\!\geq\! 3$).
 \end{thm}
    
 There is a very natural subset of $\mathcal{L}_{\bormone}$, namely the set of finitely generated $\om$-powers. If we set 
${\bf\Gamma}_f\! :=\!\{ L^\infty\mid L\mbox{ is finite}\}$, then this is $\mathcal{L}_{{\bf\Gamma}_f}$. We can decompose 
${\bf\Gamma}_f$ with respect to the cardinality, setting, for $p\!\in\!\omega$, 
${\bf\Gamma}_p\! :=\!\{ L^\infty\mid\mbox{Cardinality}(L)\! =\! p\}$, so that ${\bf\Gamma}_f\! =\!\bigcup_{p\in\om}~{\bf\Gamma}_p$. Note that ${\bf\Gamma}_0\! =\!\mathcal{L}_{\{\emptyset\}}$, and we can prove that ${\bf\Gamma}_1$ is $\bormone$-complete (see Proposition 6 in \cite{Lecomte05}). The complexity of ${\bf\Gamma}_2$ is very surprising since it is not clear at all on its definition (see Corollary 10 in \cite{Lecomte05}).

\begin{thm} \label{diff} The set ${\bf\Gamma}_2$ is $\check D_\om (\boraone )$-complete.\end{thm}

\section{$\!\!\!\!\!\!$   Open questions }\indent

 It is still open to determine all the infinite Borel ranks of the $\om$-powers of context-free languages. However  the results of 
\cite{Fin-mscs06}  suggest that the $\om$-powers of context-free languages or even of languages accepted by one-counter automata
exhibit  also a great topological complexity. Indeed, there are $\om$-languages accepted by B\"uchi one-counter automata of every Borel rank (and even of every Wadge degree) of an effective analytic set.\medskip 

In particular, for each recursive ordinal $\xi\! <\!\omega_1^{\text{CK}}$, there are some $\om$-languages $P_\xi$ and $S_\xi$  in the class   $\Borel$  such that $P_\xi$ is $\bormxi$-complete and $S_\xi$ is ${\bf \Sigma_\xi^0}$-complete. But effective analytic sets are much more complicated than $\Borel$ sets: Kechris, Marker and Sami proved in \cite{KMS89} that the supremum 
of the set of Borel ranks of   (effective)  $\Ana$ sets is the ordinal $\gamma_2^1$. 
 This ordinal is proved to be  strictly greater than the ordinal $\delta_2^1$ which is the first non ${\it\Delta}_2^1$ ordinal. 
In particular,   the ordinal $\gamma_2^1$   is   strictly greater than the ordinal $\om_1^{\text{CK}}$ (note that the exact value of the ordinal $\gamma_2^1$ may depend on axioms of  set theory).\medskip
  
 Moreover each $\om$-language $L\subseteq \Sio$ accepted by a   B\"uchi one-counter automaton is of the form 
$L = \bigcup_{1\leq j \leq n} U_j\cdot V_j^\infty$, for some one-counter finitary languages $U_j$ and $V_j$, $1\leq j \leq n$.\medskip 
 
 Therefore it seems plausible that there exist complete $\omega$-powers of a one-counter language, for each Borel class of recursive rank, and we can even 
 conjecture  that there exist some $\om$-powers of  languages accepted by  one-counter automata which have 
Borel ranks up to the ordinal  $\gamma_2^1$, although these languages are located at the very low level in the complexity hierarchy of finitary languages.

\end{document}